\documentclass[a4paper,10pt,twoside]{cpc-hepnp}
\usepackage{CJK,upgreek,fancyhdr}
\usepackage{multicol}
\usepackage{graphicx}
\usepackage{booktabs}
\usepackage{amssymb,bm,mathrsfs,bbm,amscd}
\usepackage[tbtags]{amsmath}
\usepackage{lastpage}

\begin{document}
\begin{CJK*}{GB}{gbsn}

\fancyhead[c]{\small Chinese Physics C~~~Vol. xx, No. x (201x) xxxxxx}
\fancyfoot[C]{\small 010201-\thepage}

\title{Quark-hadron phase transition in DGP including BD brane}

\author{%
      Tayeb Golanbari$^{1}$,\email{t.golanbari@uok.ac.ir; t.golanbari@@gmail.com}%
\;    Terife Haddad$^{2}$,\email{Terife$_{-}$993@yahoo.com}
\;    Abolhassan Mohammadi$^{1)}$,\email{a.mohammadi@uok.ac.ir; abolhassanm@gmail.com}
\;    M. A. Rasheed$^{3,4)}$\email{mariwan.rasheed@uhd.edu.iq}
\;    Kh. Saaidi$^{1}$,\email{ksaaidi@uok.ac.ir; khaledsaedi@gmail.com}
}

\maketitle

\address{%
$^1$ Department of Physics, Faculty of Science, University of Kurdistan, Pasdaran St. P.O. Box 66177-15175, Sanandaj, Iran.\\
$^2$ Department of Physics, Faculty of Science, Razi University, Kermanshah 6714967346, Iran.\\
$^3$ Development Center for Research and Training (DCRT), University of Human Development, Sulaimani, Kurdistan Region, Iraq. \\
$^4$ Department of Physics, College of Science, University of Sulaimani, Kurdistan Region, Iraq.\\
}

\begin{abstract}
A DGP brane-world model with a perfect fluid brane matter including a Brans-Dicke (BD) scalar field on brane has been utilized to investigate the problem of the quark-hadron phase (QHP) transition in early times of the Universe evolution. The presence of the BD scalar field comes up with some modification terms in the Friedmann equation. Since the behavior of phase transition strongly depends on the basic evolution equations, even a small change in these relations might come to interesting results about the time of transition. The phase transition is investigated using two scenarios of the first-order phase transition and smooth crossover phase transition. For first-order scenario, which is used for intermediate temperature regime, the evolution of the physical quantities, such as temperature and scale factor, are investigated before, during and after the phase transition. The results show that the transition occurs in about micro-second. In the next part, the phenomenon is studied by assuming a smooth crossover transition where the lattice QCD data is utilized to obtain a realistic equation of state for the matter. The investigation for this part is performed in two regimes of high and low-temperature. Using trace anomaly in the high-temperature regime specifies a simple equation of state which states that the quark-gluon behaves like radiation. However, in the low-temperature regime, the trace anomaly is affected by discretization effects, and the hadron resonance gas model is utilized instead. Using this model, a more realistic equation of state could be found in the low-temperature regime. The crossover phase transition in both regimes is considered. The results determine that the transition occurs at the time around a few micro-second. Also, it is realized that the transition in the low-temperature regime occurs after the transition in the high-temperature regime.

\end{abstract}

\begin{keyword}
DGP brane-world, quark-hadron phase, lattic QCD
\end{keyword}

\begin{pacs}
04.50.-h, 12.60.RC, 12.39.Hg
\end{pacs}


\begin{multicols}{2}

\section{Introduction}
According to the standard model of cosmology, as the universe expands and cools down, the universe undergoes a series of phase transitions which could lead to the generation of topological defects. In this regard, the quark-hadron phase transition could be addressed as one of the main evolution steps of the universe, so that the quark-gluon is transformed into the hadron gas. This phase transition which happens in the early universe has been studied for more than three decades and many works could be found in this regard e.g. \cite{Olive,Suhonen,Crawford,Kolb,Schramm,qcd11,Konyukhov,qcd12}. The phase transition is a specified prediction of quantum chromodynamics (QCD) that could be in general first or second-order. Also, the transition may be a smooth crossover with rapid changes that strongly depend on the values of the quark masses. Lattice QCD computation applied for two quark flavors brings up this idea that QCD makes a smooth crossover transition \cite{qcd31,qcd32,qcd33}. It is possible that in early times, the relic quark-gluon objects be created due to such a transition. \\
Mass and flavor of the quarks have a crucial role in determining the order of the phase transition. Ref.\cite{13} could be named as one of the first studies of the subject that investigates the QHP transition in the expanding universe. As the temperature of the deconfined quark-gluon plasma falls below the critical temperature $T_c$, which is about $\simeq 150 {\rm MeV}$, it is energetically capable to form color-confined hadrons in which the main part of these hadrons are pion and there are also a few number of neutrons and protons which are a direct result of the conservation of the net baryon number. As a consequence of the first-order phase transition, the hadron phase does not form immediately. As a matter of fact, to overcome the energy cost of formation of the surface of the bubble and the new hadron phase it is required to have some supercooling first. A latent heat is released as the nucleation of the hadron bubble which leads to appearance of a spherical shocking wave that crosses the surrounding supercooled quark-hadron plasma. This phenomenon warms the plasma up to the critical temperature and precludes any other nucleation in the region by one or more shock fronts. In general, bubble growth is described by deflagrations, with a shock front preceding the actual transition front. As the universe temperature reaches the critical temperature, the nucleation stops. Then, with the cost of the quark phase, the hadron bubbles grow and ultimately they percolate or coalesce. By transforming all quark-gluon phase to hadron, the transition ceases, of course, the possible quark nugget production has been ignored here \cite{14,14b,14c,14d,14e,14f,14g,14h,14i,14j,14k,14l,14m}. \\

General theory of relativity, which provides a geometrical description of spacetime, is one of the most successful theories that has passed many experimental tests and is in great agreement with the observational data. Besides the general theory of relativity, there are other alternative theories of gravity. Scalar-tensor theory could be named as one of these theories which is introduced by P. Jordan, where a four-dimensional curved manifold is embedded in five-dimensional flat spacetime \cite{Jordan}. Amongst all types of scalar-tensor theories (STT), Brans-Dicke (BD) theory is known as the most famous one \cite{BD}. BD theory, which is based on the idea of Mach, has a non-minimal coupling to gravity. The theory has great potential in solving some problems of cosmology \cite{BDhelp01,BDhelp02,BDhelp02a,BDhelp02b,BDhelp02c,BDhelp02d,BDhelp02e,BDhelp02f,BDhelp02g}.  In addition, the BD formalism of gravity also proves its abilities during the chain inflation which could facilitate the nucleation between different vacua. In \cite{Ashoorioon}, it has been shown that one of the solution of overcoming the problem of slow nucleation rate in chain inflation is to extend the gravity sector of the model to the BD theory of gravity.  \\
Another alternative is higher-dimensional theories that have a long history. The start point of these theories could be addressed to the theory of Kaluza-Klein, which a classic unified theory of electromagnetism and gravitation by utilizing the idea of fifth dimension. In 1999, inspired by M-theory, Randall and Sundrum presented a new interesting especial case of brane world scenario \cite{RS}, where the standard matter and its interaction are confined to a four-dimensional hypersurface (the brane) embedded in five-dimensional spacetime (the bulk). The extra dimension in RS brane gravity is not required to be compact. The dynamical equation governing the evolution of the brane has some modified terms in comparison to the standard cosmology such as the appearance of the quadratic term of brane energy density. However, at low energy densities, the evolution equations could come back to the standard one \cite{Binetruy-a,Binetruy-b,Shiromizu,Bogdanos,Langlois,Davis-a,Saaidi-a,Saaidi-b}. Another model of the brane-world was introduced by Dvali, Gabadadze, and Porrati (DGP) \cite{6} in 2000. The main difference of this model with RS-brane-world is the presence of Einstein term in brane action as well as the bulk action. Since the matter confined on brane is coupled with bulk gravity, some quantum corrections are induced on brane action. The curvature scalar in brane action is the result of such a correction. It is well-known that the DGP model contains two branches related to two value of $\epsilon$ parameter, which appears in Friedmann equation of the model, as $\epsilon=-1$ (normal branch) and $\epsilon=1$ (self-accelerating branch). It seems that the self-accelerating branch has a ghost-instability problem, however, this instability problem occurs at the quantum level, where the theory of gravity is unclear as well. Due to this fact, many cosmologists believe that leaving the self-accelerating branch because of its ghost-instability problem might be a rash decision in which many works about the various topics of this branch could be found \cite{14a,14aa,14ab,14ac,14ad}. \\
It is well known that the DGP model is very successful in the investigation of late time evolution of the Universe. Therefore, we are going to consider this framework for the early times evolution of the universe. The main purpose of this work is considering QHP transition in the DGP model of gravity including an STT in brane action instead of usual Einstein-Hilbert action by supposing that the phase transition is described by both the first-order phase transition and the crossover scenarios. As mentioned before, BD theory is an interesting alternative theory of gravity which overcomes some problems of general relativity in standard four-dimension cosmology. The evolution equation in this model comes into a different form with standard cosmology, and the different result is expected. \\
This paper is planed as follows: In Sec. 2, we derive the basic equations of the model. In Sec. 3, the first-order phase transition is briefly explained, then we investigate this type of phase transition in our model. Considering the QHP transition by assuming the smooth crossover approach is studied in Sec. 4, where the evolution of temperature for both high and low-temperature regime is explored. The results are summarized in Sec. 5, and some comparison with the previous results of other works is presented.\\

\section{General Framework}
In brane-world scenario, we take a four-dimensional spacetime (the brane) as the Universe which has been embedded in five-dimensional spacetime (the bulk). We consider the DGP model of  brane world scenario with the following action
\begin{eqnarray}\label{II.01}
{\cal{S}}&=& S_{bulk} + S_{brane}  \\
  & = & \int d^5x \sqrt{-\mathcal{G}} \left( {\mathcal{R} \over 2\kappa_5^2} - \Lambda \right) \nonumber \\
  & + & \int d^4x\sqrt{-g}\left( [\tilde{K}] + {1 \over 2} \Big[ \phi
R-{\omega\over{\phi}}\phi^{,\alpha}\phi_{,\alpha} \Big] +{\cal{L}}_m \right), \nonumber
\end{eqnarray}
where $ \tilde{K} = {[K] /\kappa_5^2} $ and $ [K] $ is exterior curvature. The action is presented in term of five-dimensional coordinates $(x^0, x^1, x^2, x^3, y)$ on bulk, where the hypersurface $y=0$ describes the brane. In the bulk action, $\mathcal{G}$ is the determinant of five-dimensional metric $\mathcal{G}_{AB}$  with signature $(- + + + +)$, $\mathcal{R}$ is five-dimensional Ricci scalar constructed from $\mathcal{G}_{AB}$, $\Lambda$ is five-dimensional cosmological constant, and $\kappa_5$ is a redefinition of five dimensional Newtonian gravitational constant $G_5$ as $\kappa_5^2= 8\pi G_5$, in which is related to five dimensional Plank mass by $\kappa_5^2=M_5^{-3}$. In the brane action, $ g $ and $ R $ are determinant of four-dimensional induced metric and Ricci scalar related to induced metric
$g_{\mu\nu}=\delta^A_{\ \mu} \delta^B_{\ \nu} \mathcal{G}_{AB}$ respectively. The BD scalar field is denoted by $\phi$ which lives on brane without potential term, and $ \omega $ is a dimensionless coupling constant which determines the coupling between gravity and $\phi$. {\bf The Lagrangian term $\mathcal{L}_m$ in the brane action actually includes two part as the Lagrangian of the brane-matter, denoted by $L_m$, and brane tension, $\lambda$. Then it could be write as $\mathcal{L}_m= L_m - \lambda$.} \\

The Einstein field equation is derived by taking variation of action with respect to metric as
\begin{equation}\label{II.02}
^{(5)}G_{AB} = \mathcal{R}_{AB} - \mathcal{G}_{AB}\mathcal{R} = \kappa_5^2 \left[ T^{(\Lambda)}_{AB} + T_{AB}\delta(y) \right].
\end{equation}
Also, the energy-momentum tensor of the bulk cosmological constant is described by $T^{(\Lambda)}_{AB}=-\Lambda \mathcal{G}_{AB}$, and $T_{AB}$ is the total brane energy-momentum tensor with the following definition
\begin{eqnarray}\label{II.03}
T_{AB} & = & g_A^{\ \ \mu}g_B^{\ \ \nu} \Big[ \phi T^{G}_{\mu\nu} + T^{(\phi)}_{\mu\nu} + T^{(m)}_{\mu\nu}  \Big],
\end{eqnarray}
in which $T^G_{\mu\nu}$ is due to presence of curvature scalar in brane action and behave as a source of gravity. $T^{(\phi)}_{\mu\nu}$ is related to BD scalar field and $T^{(m)}_{\mu\nu}$ is related to the brane-matter energy-momentum tensor. The matter section of brane is assumed to be filled with perfect fluid $T^{(m)\mu}_{\ \ \ \nu} = \delta(y) {\rm diag}(-\rho_b, p_b, p_b, p_b, 0)$, where $\rho_b=\rho + \lambda$ is the brane-matter energy density, and $p_b=p-\lambda$ is the brane-matter pressure. \\

Since brane matter has no direct coupling to the other component in the action, the continuity of brane matter is preserved, i.e.
\begin{equation}\label{conservationEq}
\dot\rho + 3H(\rho+p)=0.
\end{equation}
On the other hand, the equation of motion of BD scalar field $\phi$ is obtained by taking variation of action with respect to scalar field as
\begin{equation}\label{II.05}
{2\omega \over \phi}- \triangle \phi - {\omega \over \phi^2} \partial_\alpha \phi \partial^\alpha \phi + R = 0 .
\end{equation}
Following \cite{Binetruy-a,Binetruy-b,Shiromizu,8,9}, the Friedmann equation of the model could be derived after some algebraic analysis as

\begin{equation}\label{II.24}
H^2 ={\Lambda \over 6 M_5^3} + \frac{1}{36 M_5^6} \Bigg[ 3A\phi H^2  + (\rho+\lambda) \Bigg]^2 + {\zeta \over a^4}
\end{equation}
where $A=  {\beta^2\omega / 6} -(1 +\beta) $; note that in obtaining above equation it is assumed that $\phi = \phi_0 a^\beta$ \cite{15,15a}. The effective DGP length scale in this model could be defined as $ l_{effDGP} = {|A|\phi / M_5^3}$,
where $\phi$ is a function of time, then the effective DGP length scale is varying by time. One could introduced a constant DGP length scale as $l_c = {|A| \phi_0 / M_5^3}$. Then, the familiar form of Friedmann equation is given by
\begin{eqnarray}\label{II.26}
H^2  = {2 \over \Phi^2} \left(\chi + \epsilon \sqrt{ \chi^2 - \Phi^2 \left[ \Lambda_4 + {\rho_b^2 - \lambda^2 \over 36 M_5^6}  \right]}\right),
\end{eqnarray}
where $ \chi=\big[1- (\Phi \rho_b) / 6 M_5^3 \big] $, $ \Phi=A \phi / M_5^3 = A \Phi_0 a^\beta$ (with $\Phi_0 \equiv \phi_0 / M_5^3$) and $\epsilon = \pm 1$. Also the brane effective cosmological constant is described
by $\Lambda_4 = \big[\Lambda + ( \lambda^2 / 6 M_5^3) \big] / 6 M_5^3$, which same as RS-model has been assumed to be zero. The last term which describes the dark radiation has been ignored as well. {\bf Following \cite{25,27a}, for the convenience, the five-dimensional Planck mass is set to one, i.e. $\kappa_5=1$, for the rest of the work.} \\

\section{First order quark hadron phase transition}
The phase transition in the temperature interval $180 {\rm MeV} < T < 250 {\rm MeV}$ is characterized by sigular behavior of the partition function which might be first or second order phase transition \cite{23}. The phase transition in higher or lower temperature occurs through trace anomaly or hadronic resonance gas which will be investigated in next section. In this section, we assumed that the quark-hadron phase transition is first order which happen in aforementioned temperature interval (the readers should refer to \cite{24} and references therein for more information about first order quark-hadron phase transition). Following \cite{23}, the matter equation of state in quark-gluon phase could be stated as
\begin{equation}\label{IV.39}
\rho _q=3a_q T^4 +W(T), \qquad p_q =a_q T^4 -W(T),
\end{equation}
where the subscript $q$ stands for quark-gluon matter and the constant $a_q$ is equal to $a_q=61.75(\pi^2/90)$. $W(T)$ which denotes the potential energy density is given by \cite{23},
\begin{equation}\label{IV.40}
W(T)=B+\gamma_T T^2 - \alpha_{T} T^{4},
\end{equation}
in which the constant $B$ is the bag pressure constant, $\alpha_T = 7\pi^2/20$, $\gamma_T = m^2_s/4$ where the mass of the strange quark, $m_s$, stands in the range $(60 - 200)$ MeV. This type of the potential $W$ could be found out in the model in which the quark fields interact with a chiral formed from a scalar field and the $\pi$ meson field \cite{davis2000}. The constant $B^{1/4}$ in above equation could be determined through the results obtained from low energy hadron spectroscopy, heavy ion collisions, and  from phenomenological fits of light hadron properties stating that it should be between 100 and 200 MeV \cite{24}. \\
The cosmological fluid which commonly is taken as an idea gas, could be described by Maxwell-Boltzmann distribution function with energy density $\rho_h$ and pressure $p_h$ obeying the following equation of state
\begin{equation}\label{IV.41}
p_h = {1 \over 3} \rho_h = a_\pi T^{4}.
\end{equation}
where the constant is read as $a_\pi = 17.25 \pi^2 / 90$. On the other hand, it has been shown that the identity $p_q(T_c)=p_h(T_c)$ defines a critical temperature $T_c$ \cite{12,12a,12b,12c,12d} in which for the present model one has
\begin{equation}\label{IV.42}
 T_c={\Bigg[{\frac{\gamma_T + \sqrt{\gamma^2_T+ 4B(a_q+\alpha_T-a_{\pi})}}{2(a_q+\alpha_T -a_{\pi})}}\Bigg]}^{1/ 2}\approx 125\   {\rm MeV}.
\end{equation}
where in obtaining the above result it is assumed that $m_s = B^{1/4} = 200 {\rm MeV}$. \\
Because of this fact that the phase transition is first order, {\bf there could be found a discontinuities in energy density, number density and entropy, when they are crossing the critical curve. This behavior of the quantities are perfectly described in \cite{14}.} The main quantities that we are dealing with them in the present work are energy density, pressure, scale factor and the temperature which are determined through the conservation equation, Friedmann equation and equation of state. In the subsequent subsections, the behavior of these quantities will be studies in the framework of DGP brane-world including a BD scalar field in the brane before, during and after the phase transition. \\

\subsection{Behavior of temperature in the quark-gluon phase (QGP) for general $ W(T) $}
In the temperature higher than the critical temperature $T > T_c$, i.e. before the phase transition, the matter is in the quark phase which is described by the equation of state Eq.(\ref{IV.39}). The Hubble parameter is determined by applying the equation of state Eq.(\ref{IV.39}) on the conservation equation (\ref{conservationEq}), as
\begin{equation}\label{IV.43}
H = - \Big[ {3a_q-\alpha_T \over 3a_q} + {\gamma_T \over 6a_qT^2}\Big] {\dot{T} \over T}.
\end{equation}
where the potential energy density (\ref{IV.40}) is utilized. By integrating from the above equation, one could extract the scale factor of the universe in this phase as
\begin{equation}\label{IV.44}
a(T) = c T^{{\alpha_T - 3a_q \over 3a_q}} \exp\Big( {\gamma_T \over 12a_q T^2} \Big).
\end{equation}
Finally, the evolution of the temperature is obtained by substituting the above results in the Friedmann equation (\ref{II.26}), that is read as
\begin{eqnarray}\label{IV.45}
\dot{T} & = &  - {6a_qT^3 \over 2(3a_q - \alpha_T)T^2 + \gamma_T} \\
 &  & \times \Bigg[  {2 \over \Phi^2} \left(\chi + \epsilon \sqrt{ \chi^2 - {\Phi^2 \over 36} (\rho_b^2 - \lambda^2)}\right)\Bigg]^{1/2} \nonumber .
\end{eqnarray}
which describe the behavior of the temperature in terms of the cosmic time. The obtained numerical result for the temperature is depicted in Fig.\ref{fig3a} and \ref{fig3b} respectively for normal branch ($\epsilon=-1$) and self-accelerating branch ($\epsilon=+1$). It is illustrated that the temperature in quark phase decreases by passing time in which the increasing of the BD coupling $\omega$ results in the reduction of temperature. Also, for the self-accelerating branch of the model, the temperature reaches the critical temperature faster than the normal branch.

\begin{center}
\includegraphics[width=8cm]{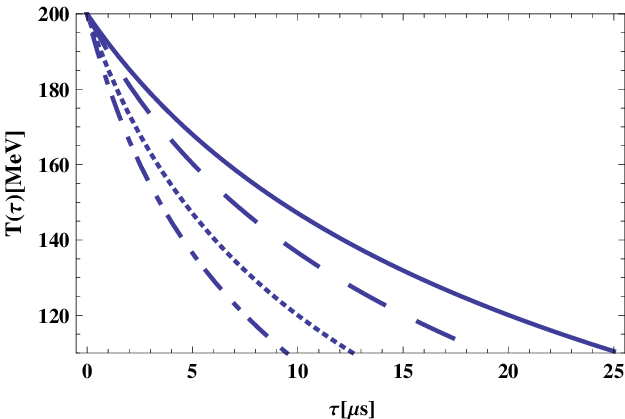}
\figcaption{\label{fig3a}\label{fig3ab} $\mathbf{\underline{\epsilon=-1}}$: $T$ versus $\tau$ in the quark-gluon phase has been plotted for different
values of BD coupling constant $\omega$ as: $\omega = 2.1 \times 10^{3}$ (solid line),
$\omega = 2.3 \times 10^{3}$ (dashed line), $\omega = 2.5 \times 10^{3}$ (dotted line),
$\omega = 2.7 \times 10^{3}$ (dotted-dashed line). The other constant parameters are
taken as: $\beta=5.24 \times 10^{-2}$, $\Phi_0=2 \times 10^4 {\rm MeV^{-1}}$,
$\lambda= 10^{9} {\rm MeV^4}$ and $B^{1/4}=200 {\rm MeV}$.}
\end{center}
\begin{center}
\includegraphics[width=8cm]{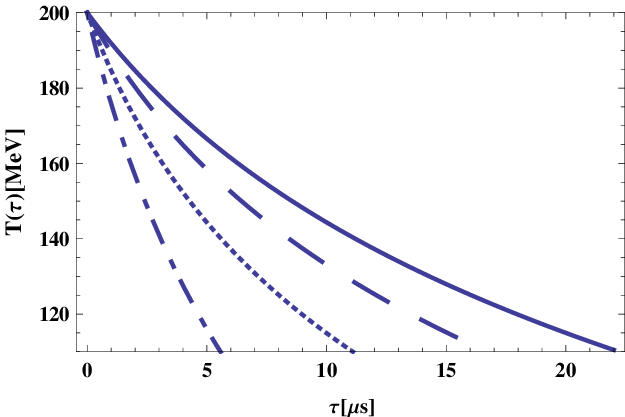}
\figcaption{\label{fig3b}\label{fig3ab} $\mathbf{\underline{\epsilon=+1}}$: $T$ versus $\tau$ in the quark-gluon phase has been plotted for different
values of BD coupling constant $\omega$ as: $\omega = 2.1 \times 10^{3}$ (solid line),
$\omega = 2.3 \times 10^{3}$ (dashed line), $\omega = 2.5 \times 10^{3}$ (dotted line),
$\omega = 2.7 \times 10^{3}$ (dotted-dashed line). The other constant parameters are
taken as: $\beta=5.24 \times 10^{-2}$, $\Phi_0=2 \times 10^4 {\rm MeV^{-1}}$,
$\lambda= 10^{9} {\rm MeV^4}$ and $B^{1/4}=200 {\rm MeV}$.}
\end{center}

\subsubsection{Evolution of temperature in the QGP for $ W(T)=B $}
Ignoring the effect of the temperature results in a constant potential $W(T)=B$ which is another interesting model that deals with quark confinement and worth being considered. This is related to an elastic bag model which allows the quarks to freely move around. Due to this constant potential, the equation of state gets a simple form $p_q = {(\rho_q -4B) / 3}$, leading to the following Hubble parameter and scale factor
\begin{equation}\label{IV.46}
H = - {\dot{T} \over T}  ; \qquad a(T) = {c \over T}.
\end{equation}
Applying above result Eq.(\ref{IV.46}) in the Friedmann equation, comes to the temperature evolution equation which is given by
\begin{equation}\label{IV.47}
\dot{T} =  - T
\Bigg[ {2 \over \Phi^2} \left(\chi + \epsilon \sqrt{ \chi^2 - {\Phi^2 \over 36} (\rho_b^2 - \lambda^2)}\right)\Bigg]^{1/2}
\end{equation}
The equation gives behavior of the temperature in terms of the cosmic time. Solving the above differential equation numerically, the result is plotted in Figs.(\ref{fig4a}) and (\ref{fig4b}) respectively for normal and self-accelerating branches for different choices of the BD coupling $\omega$, where one could find that the enhancement of BD constant results in lower temperature. In general, the temperature decreases by passing time and it happens faster for self-accelerating branch than the normal branch. \\

\begin{center}
\includegraphics[width=8cm]{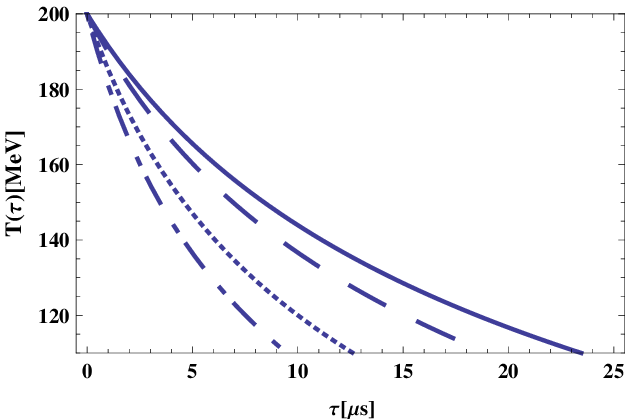}
\figcaption{\label{fig4a}\label{fig4ab} $\mathbf{\underline{\epsilon=-1}}$: Temperature $T$, as a function of time, $t$, in the quark-gluon phase
has been plotted for constant self-interacting potential $W(T)=B$, and for different
values of BD coupling constant $\omega$ as: $\omega = 2.1 \times 10^{3}$ (solid line),
$\omega = 2.3 \times 10^{3}$ (dashed line), $\omega = 2.5 \times 10^{3}$ (dotted line),
$\omega = 2.7 \times 10^{3}$ (dotted-dashed line). The other constant parameters are
taken as: $\beta=5.24 \times 10^{-2}$, $\Phi_0=2 \times 10^4 {\rm MeV^{-1}}$,
$\lambda= 10^{9} {\rm MeV^4}$ and $B^{1/4}=200 {\rm MeV}$.}
\end{center}
\begin{center}
\includegraphics[width=8cm]{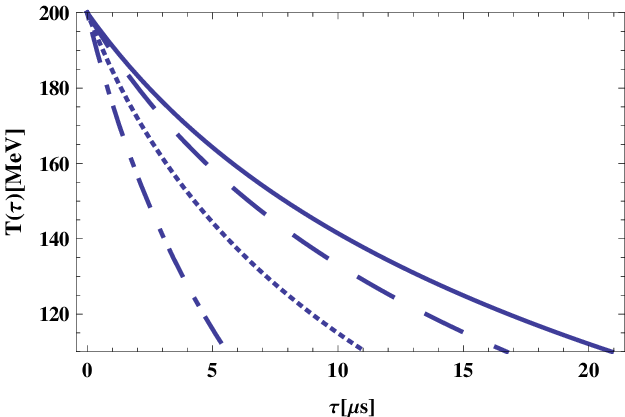}
\figcaption{\label{fig4b}\label{fig4ab} $\mathbf{\underline{\epsilon=+1}}$: Temperature $T$, as a function of time, $t$, in the quark-gluon phase
has been plotted for constant self-interacting potential $W(T)=B$, and for different
values of BD coupling constant $\omega$ as: $\omega = 2.1 \times 10^{3}$ (solid line),
$\omega = 2.3 \times 10^{3}$ (dashed line), $\omega = 2.5 \times 10^{3}$ (dotted line),
$\omega = 2.7 \times 10^{3}$ (dotted-dashed line). The other constant parameters are
taken as: $\beta=5.24 \times 10^{-2}$, $\Phi_0=2 \times 10^4 {\rm MeV^{-1}}$,
$\lambda= 10^{9} {\rm MeV^4}$ and $B^{1/4}=200 {\rm MeV}$.}
\end{center}

\subsection{Behavior of the hadron volume fraction}
During the phase transition the matter density changes from the quark density $\rho_Q$ to the hadron energy density $\rho_H$ where they are respectively given by $\rho_Q = 5 \times 10^9 $ MeV$^4$, $\rho_H \approx 1.38 \times 10^9 $ MeV$^4$ (Note that $\rho_Q$ stands for the energy density of the universe when it is in pure quark phase, namely when all matter of the universe is quark. Similarly $\rho_H$ stands for the energy density of the universe when it is in pure hadron phase). In this phase transition the temperature, pressure, entropy and enthalpy remains constant in which $T_c=125$ MeV, and constant pressure $p_c \approx 4.6 \times 10^8 $ MeV$^4$. \\
The universe transits from a pure quark phase to a pure hadron phase. In the middle of the transition, the energy density of the universe is a composite of quark and hadron. Then, the energy density could be expressed in terms of a volume fraction term in hadron phase which is defined as \cite{24, 25, 27,27a,27aa}
\begin{equation}\label{IV.48}
\rho = \rho_H h(t) + [1-h(t)]\rho_{Q},
\end{equation}
It is realized that at the beginning of the transition ($h(t)=0$) all matters are in the phase of quark, and by passing time the matter phase changes to the hadron phase in which at the end of the transition ($h(t)=1$) all matters are in the phase of hadron. Imposing Eq.(\ref{IV.48}) on the conservation equation comes to the following Hubble parameter
\begin{equation}\label{IV.49}
H=-{1 \over 3} {(\rho_H - \rho_Q)\dot{h} \over \rho_Q+p_c+(\rho_H-\rho_Q)h}
= -{1 \over 3} {r\dot{h} \over 1+rh},
\end{equation}
where the parameter $r$ is defined as $r=(\rho_H - \rho_Q) / (\rho_Q+p_c)$. The scale factor is obtained by taking integrate from the above relation as
\begin{equation}\label{IV.50}
a(t) = a_0 \big[ 1+rh \big]^{-1/3}.
\end{equation}
Fig.(\ref{fig5ah}) illustrates the scale factor of the universe versus the volume fraction during the quark hadron phase transition which expresses an expansion for the universe in the era where the temperature, pressure, entropy and enthalpy are constant. \\
\begin{center}
\includegraphics[width=8cm]{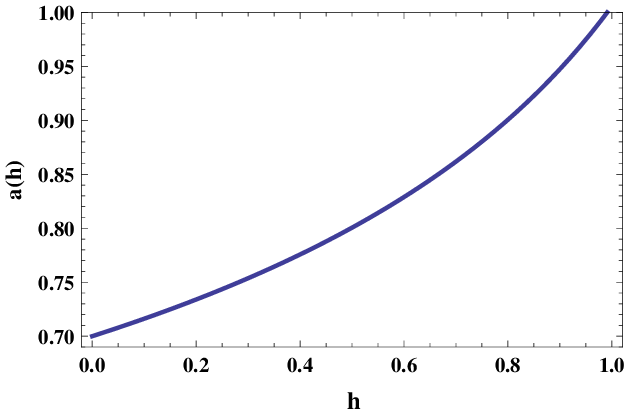}
\figcaption{\label{fig5ah} Scale factor as a function of the hadron volume fraction
during the QHPT in the DGP brane gravity with BD scalar field in brane.}
\end{center}
By Utilizing the Hubble parameter (\ref{IV.49}) in the Friedmann equation (\ref{II.26}), one could find the evolution equation of the volume fraction as
\begin{eqnarray}\label{IV.51}
\dot{h} & = & - {3(1+rh) \over r} \\
 & & \times  \Bigg[  {2 \over \Phi^2} \left(\chi + \epsilon \sqrt{ \chi^2 - {\Phi^2 \over 36} (\rho_b^2 - \lambda^2)}\right)  \Bigg]^{1/2}. \nonumber
\end{eqnarray}
By finding the numerical solution for the above differential equation, the behavior of the volume fraction in terms of the cosmic time during the phase transition is presented in Fig.(\ref{fig6a}) and (\ref{fig6b}) for different choices of the BD coupling $\omega$ for both branches. Bigger BD coupling results in faster phase transition. Also, for the self-accelerating branch, the transition happens earlier and the universe enters to the pure hadron phase faster than the normal branch.
\begin{center}
\includegraphics[width=8cm]{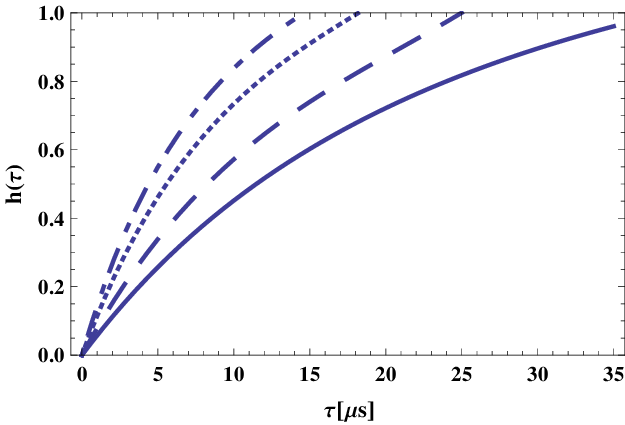}
\figcaption{\label{fig6a}\label{fig6ab} $\mathbf{\underline{\epsilon=-1}}$: Hadron volume fraction as a function of  cosmic time $\tau$
has been depicted for different values of BD coupling constant $\omega$ as:
$\omega = 2.1 \times 10^{3}$ (solid line), $\omega = 2.3 \times 10^{3}$ (dashed line),
$\omega = 2.5 \times 10^{3}$ (dotted line), $\omega = 2.7 \times 10^{3}$
(dotted-dashed line). The other constant parameters are taken as:
$\beta=5.24 \times 10^{-2}$, $\Phi_0=2 \times 10^4 {\rm MeV^{-1}}$,
$\lambda= 10^{9} {\rm MeV^4}$ and $B^{1/4}=200 {\rm MeV}$.}
\end{center}
\begin{center}
\includegraphics[width=8cm]{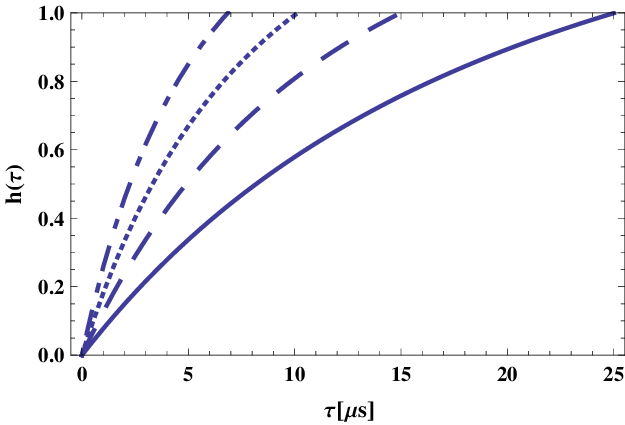}
\figcaption{\label{fig6b}\label{fig6ab} $\mathbf{\underline{\epsilon=+1}}$: Hadron volume fraction as a function of  cosmic time $\tau$
has been depicted for different values of BD coupling constant $\omega$ as:
$\omega = 2.1 \times 10^{3}$ (solid line), $\omega = 2.3 \times 10^{3}$ (dashed line),
$\omega = 2.5 \times 10^{3}$ (dotted line), $\omega = 2.7 \times 10^{3}$
(dotted-dashed line). The other constant parameters are taken as:
$\beta=5.24 \times 10^{-2}$, $\Phi_0=2 \times 10^4 {\rm MeV^{-1}}$,
$\lambda= 10^{9} {\rm MeV^4}$ and $B^{1/4}=200 {\rm MeV}$.}
\end{center}

\subsection{Evolution of temperature in Pure hadronic phase era}
After the phase transition, the universe stands in the pure hadronic phase where the following simple equation describes the equation of state of the matter
\begin{equation}\label{IV.52}
\rho_h=3p_h=3a_\pi T^{4}.
\end{equation}
Then, following the same processes as before, the Hubble parameter and the scale factor are easily derived as
\begin{equation}\label{IV.53}
H = -{\dot{T} \over T} ; \qquad a(T) =c {T_c \over T}.
\end{equation}
Inserting above equation in the Firedmann equation results in a evolution equation for the temperature in the pure hadronic phase of the universe
\begin{equation}\label{IV.54}
\dot{T} =  - T
 \Bigg[ {2 \over \Phi^2} \left(\chi + \epsilon \sqrt{ \chi^2 - {\Phi^2 \over 36} (\rho_b^2 - \lambda^2)}\right)\Bigg]^{1/2}.
\end{equation}
Solving above differential equation numerically, the behavior of the temperature in terms of the cosmic time could be depicted as Figs.\ref{fig7a} and \ref{fig7b} respectively for the normal and self-accelerating branches for different {\bf values BD coupling constant $\omega$}. It is clearly seen that the temperature decreases by passing time and the hadron phase era of the universe occurs earlier in the self-accelerating branches. Also, for higher values of $\omega$ there is lower temperature and the phase transition happens earlier. This result is in good consistency with the previous results regarding the other eras.
\begin{center}
{\includegraphics[width=8cm]{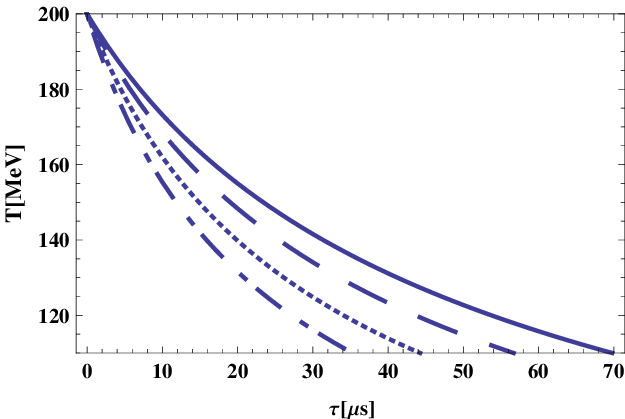}
\figcaption{\label{fig7a}\label{fig7ab} $\mathbf{\underline{\epsilon=-1}}$: Evolution of temperature $T$ versus $t$ in the pure hadronic
phase era has been depicted for different values of BD coupling constant $\omega$ as: $\omega = 2.1 \times 10^{3}$ (solid line),
$\omega = 2.3 \times 10^{3}$ (dashed line), $\omega = 2.5 \times 10^{3}$ (dotted line),
$\omega = 2.7 \times 10^{3}$ (dotted-dashed line). The other constant parameters are taken as: $\beta=5.24 \times 10^{-2}$,
$\Phi_0= 2 \times 10^4 {\rm MeV^{-1}}$ and $\lambda= 7 \times 10^{10} {\rm MeV^4}$.}}
\end{center}

\begin{center}
{\includegraphics[width=8cm]{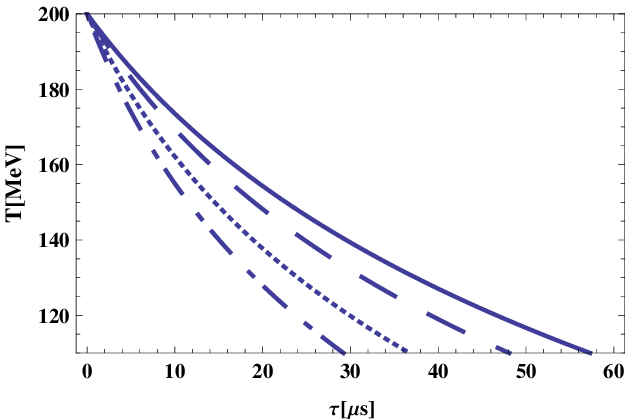}
\figcaption{\label{fig7b}\label{fig7ab} $\mathbf{\underline{\epsilon=+1}}$: Evolution of temperature $T$ versus $t$ in the pure hadronic
phase era has been depicted for different values of BD coupling constant $\omega$ as: $\omega = 2.1 \times 10^{3}$ (solid line),
$\omega = 2.3 \times 10^{3}$ (dashed line), $\omega = 2.5 \times 10^{3}$ (dotted line),
$\omega = 2.7 \times 10^{3}$ (dotted-dashed line).The other constant parameters are taken as:    $\beta=5.24 \times 10^{-2}$,
$\Phi_0= 2 \times 10^4 {\rm MeV^{-1}}$ and $\lambda= 7 \times 10^{10} {\rm MeV^4}$.}}
\end{center}

\section{Lattice QCD phase transition}
In this section, we are about to investigate the lattice QCD phase transition and consider the physical quantities related to the QHP transition in the context of DGP brane-world scenario with brane BD scalar field. A brief review on lattice QCD is presented in the following lines. One of the basic phenomena in the particle physics is lattice QCD phase transition, so that any study related to the early time evolution of the universe involves this phase transition. According to this scenario, there is a soup of quarks and gluons which interact. Then, through a crossover transition, hadrons  are formed. There are various different method to acquire the equation of state. Studying the nonperturbative regime of the QCD equation of state is possible via the newly presented approach as lattice QCD. In the recent calculation, the equation of state has been estimated for 2+1 flavor QCD. The most extensive estimation of equation of state has been performed  with fermion formulation on lattice with temporal extent $N_t=4, 6$ \cite{16,17,17a}, $N_t=8$ \cite{18} and $N_t=6, 8, 10$ \cite{19}. For high temperature region $T > 250 {\rm MeV}$, one could accurately compute the trace anomaly. Then, in high temperature regime, the lattice QCD data for the trace anomaly could be utilized  in order to construct a realistic equation of state. \\
For the low temperature regime $T < 180 {\rm MeV}$, the situation is somehow different. In this regime, the trace anomaly is affected by large discretization effect. However, the hadronic resonance gas (HRG) model is a an appropriate approach to construct an equation of state, which has been studied in \cite{20}.

\subsection{High temperature regime}
In this regime of temperature $T > 250 {\rm MeV}$, the gluon and quarks effectively are massless and behave like a radiation. Applying the trace anomaly for obtaining an equation of state determines that the data are fit on a simple form of equation of state as
\begin{equation}\label{III.29}
\rho(T) \simeq \alpha T^4, \qquad p(T) \simeq \sigma T^4 ,
\end{equation}
where the constants $\alpha$ and $\beta$ are found out using a least-squares fit as $\alpha = 14.9702 \pm 009997$ and $\sigma = 4.99115 \pm 004474$ \cite{17,17a}. Substituting Eq.(\ref{III.29}) into the conservation equation leads to following result which expresses the scale factor in terms of the temperature as
\begin{equation}\label{III.30}
a(T)=cT^{-\frac{4\alpha}{3(\alpha+\sigma)}},
\end{equation}
where $c$ is a constant of integration. Substituting Eq.(\ref{III.30}) into Eq.(\ref{II.26}), the time evolution of temperature in this regime is derived as
\begin{equation}\label{III.31}
\dot{T} =  - {3(\alpha+\sigma)T \over 4\alpha}
  \left[ {2 \over \Phi^2} \left(\chi + \epsilon \sqrt{ \chi^2 - {\Phi^2 \over 36} (\rho_b^2 - \lambda^2)}\right) \right]^{1/2}.
\end{equation}
which describes the behavior of temperature as a function of time. The numerical results for above differential equation have been depicted in Fig.\ref{fig1a} for two branches.

We have plotted the numerical results of Eq.(\ref{III.31}) in figure \ref{fig1a} for two branches. The figures display the effective temperature in the QGP in BD model of DGP brane world for temperature interval $ 250 {\rm MeV} \leq T \leq 700 {\rm MeV}$, acquired for the smooth crossover approach. By passing time, the universe gets cooler and the situation is almost the same for both cases. The higher and lower curves respectively indicate the temperature behavior for the initial temperatures $700 {\rm MeV}$ and $250 {\rm MeV}$. The curve for any other initial value between these two temperatures stands in the shaded region. \\
\end{multicols}

\begin{center}
\includegraphics[width=17cm]{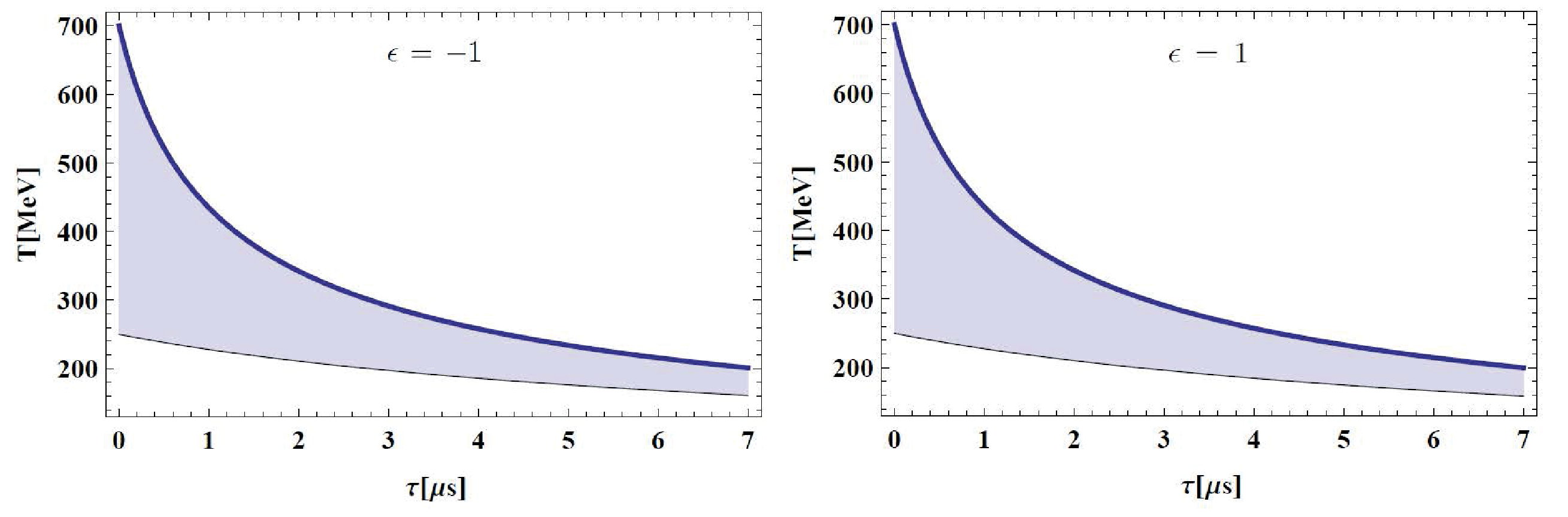}
\figcaption{\label{fig1a}\label{HT}The figure displays the behavior of the temperature $T$ versus the cosmic time $\tau$ for high temperature region of the smooth crossover procedure in the DGP brane gravity with BD scalar field
on the brane for the interval temperature $ 250 {\rm MeV} \leq T \leq 700 {\rm MeV}$. The constant parameters are taken as: $\beta=2.47 \times 10^{-3}$, $\Phi_0=2 \times 10^{5} {\rm MeV^{-1}}$, $\omega=10^4$ and $\lambda= 10^{9} {\rm MeV^4}$. }
\end{center}

\begin{multicols}{2}
\subsection{Low temperature regime}
Although the trace anomaly is known as a suitable approach to acquire a realistic equation of state in high temperature regime, the approach seems to fail in low temperature regime mostly due to discretization effect. In low temperature regime, where $T < 180 {\rm MeV}$, one can achieve an equation of state by using the HRG model \cite{20}. In HRG model, the confinement of QCD is treated as non-interacting gas of fermions and bosons \cite{21,21a}. In this case, the era before the phase transition (namely quark-gluon phase) is considered at low temperature. Then the Universe is in confinement phase and could be treated as a non-interacting gas of fermions and bosons. The idea of the HRG model is to implicitly account for the strong interaction in the confinement phase by looking at the hadronic resonances only, since these are basically the only relevant degrees of freedom in that phase. It has been shown that, the HRG results can be parametrized for trace anomaly as \cite{20}:
\begin{equation}\label{III.32}
{I(T) \over T^4}\equiv  {\rho -3p \over T^4} = a_1T + a_2T^3 + a_3T^4 + a_4T^{10},
\end{equation}
where $a_1$ = 4.654 GeV$^{-1}$, $a_2$ = -879 GeV$^{-3}$, $a_3$ = 8081 GeV$^{-4}$, $a_4$ = -7039000 GeV$^{-10}$, and $I(T) = \rho(T) - 3p(T)$ is the trace anomally. Through the calculation of trace anomally $I(T)$ in lattice QCD, the energy density, pressure and entropy could be estimated by using usual thermodynamics identities. Integral of trace anomaly displays the pressure difference between two temperature $T$ and $T_{\rm low}$
\begin{equation}\label{III.33}
{p(T) \over T^4 } - {p(T_{\rm low})\over T^4_{\rm low}} = \int^T_{T_{\rm low}} {dT' \over T'^5}I(T').
\end{equation}
The second term on the left hand side of above equation, $p \left( T_{\rm low}\right) $,  could be ignored for sufficiently small values of the lower integration limit because of exponential suppression \cite{22}. The energy density which is expressed as $\rho(T) = I(T) + 3p(T)$ could be calculated. At the end, by using Eqs.(\ref{III.32}) and (\ref{III.33}), we have following relations for energy density and pressure respectively
\begin{eqnarray}\label{III.34}
\rho(T) &=&  3\eta T^4 + 4a_1 T^5 + 2a_2 T^7 + {7a_3 \over 4}T^8 + {13a_4 \over 10} T^{14}, \qquad\\
p(T) &=& \eta T^4 + a_1 T^5 + {a_2 \over 3} T^7 + {a_3 \over 4}T^8 + {a_4 \over 10} T^{14}, \label{III.35}
\end{eqnarray}
where $\eta = -0.112$. \\
Inserting Eqs.(\ref{III.34}) and (\ref{III.35}) in the conservation equation, the Hubble parameter can be derived in terms of temperature of its time derivative as
\begin{equation}\label{III.36}
H  =  - {12\eta T^3 + 20a_1T^4 + A(T) \over 3\big[ 4\eta T^4 + 5a_1T^3 + B(T) \big]} \dot{T},
\end{equation}
where the parameters $A(T)$ and $B(T)$ are defined by
\begin{eqnarray}
A(T) & = & 14 a_{2}T^{6} + 14 a_{3}T^{7} + \frac{91}{5}a_{4} T^{13},\nonumber \\
B(T) & = &  \frac{7}{3} a_{2} T^{7} + 2a_{3} T^{8} + \frac{7}{5}a_{4}T^{14}.\nonumber
\end{eqnarray}
To obtain the scale factor as a function of temperature, we should integrate Eq.(\ref{III.36}), which results in the following expression
\begin{equation}\label{III.37}
a(T) = {c \over T\big[ 60 \eta + 75a_1T + 35a_2 T^3 + 30a_3 T^4 + 21 T^{10} \big]^{1/3}},
\end{equation}
here $c$ is constant of integration. Finally using Eq.(\ref{III.36}) and the Friedmann equation (\ref{II.26}), one can find out the time evolution of temperature
\begin{equation}\label{III.38}
\dot{T} =  - {3\big[ 4\eta T^4 + 5a_1T^3 + B(T) \big]\over 12\eta T^3 + 20a_1T^4 + A(T)}
  \left[  {2 \over \Phi^2} \left(\chi + \epsilon \sqrt{ \chi^2 - {\Phi^2 \over 36} (\rho_b^2 - \lambda^2)}\right)\right]^{1/2}.
\end{equation}
\end{multicols}

\begin{center}
\includegraphics[width=17cm]{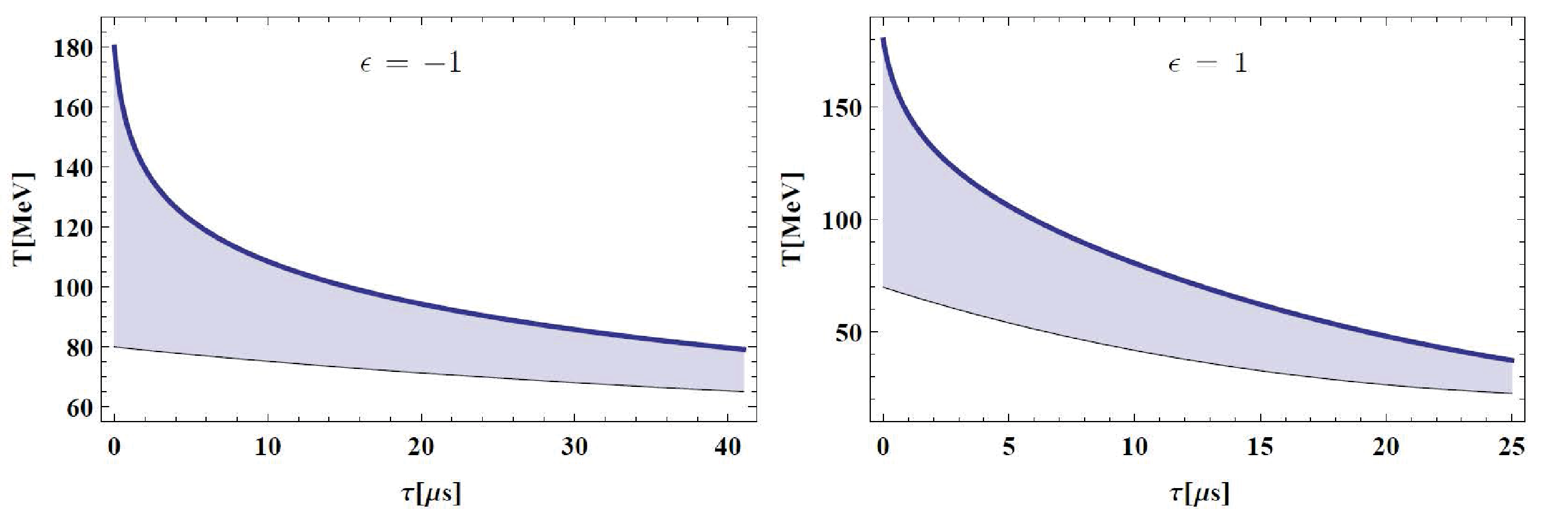}
\figcaption{\label{LT}\label{fig2a} $T$ versus $\tau$ according to the low temperature region of
the  smooth crossover procedure in the DGP brane gravity with BD scalar field
on the brane for the temperature interval $80 {\rm MeV} \leq T \leq 180 {\rm MeV}$ and following constants: $\beta=2.47 \times 10^{-3}$, $\Phi_0=2 \times 10^{5} {\rm MeV^{-1}}$, $\omega=10^4$ and $\lambda= 10^{9} {\rm MeV^4}$. }
\end{center}
\begin{multicols}{2}

Differential Eq.(\ref{III.38}) describes the behavior of temperature as a function of cosmic time in the QGP for DGP brane-world scenario with a BD scalar field on the brane for the temperature interval $80 {\rm MeV} \leq T \leq 180 {\rm MeV}$. The relation displays the era before phase transition at low temperature where the Universe is treated as a non-interacting gas of fermions and bosons since it is in confinement phase. The equation could be solved numerically in which the obtained results have been plotted in Fig.\ref{fig2a} for two branches of the model. The figure illustrates the behavior of temperature in terms of the cosmic time. The higher and lower curves display respectively the temperature behavior for the initial value $180 {\rm MeV}$ and $80 {\rm MeV}$. Also, the corresponding curve for any other initial value in this temperature interval stands in the shaded region. Also, it is vividly seen that the phase transition for normal branch occurs after the self-accelerating branch. \\

\section{Conclusion}
In this work, we have investigated the quark-hadron phase transition in the Brans-Dicke DGP model of Brane Gravity by considering two different scenarios. We studied the evolution of physical quantities relevant to the physical description of the early times such as energy density, scale factor, and temperature, for two cases of $\epsilon=-1$ and $\epsilon=+1$ which are known as normal and self-accelerating branches respectively, in the original DGP model. The quark-hadron phase transition was investigated using two scenarios as the first-order phase transition and smooth crossover phase transition. The first-order phase transition scenario works for intermediate temperature, i.e. $180 {\rm MeV} < T < 250 {\rm MeV}$. Applying this scenario, the behavior of the physical quantities, such as temperature, energy density and scale factor, were studied for different stages as: before (quark-gluon phase), during and after (hadron phase) the phase transition for both branches of the model as the normal branch and self-accelerating branch. The result shows during the phase transition the universe is expanding while the temperature and pressure are constant, and the universe leaves from a pure quark phase and enters a pure hadronic phase. The phase transition for this scenario depends on the values of the BD coupling constant in which for higher $\omega$ we have a faster transition. Also, the phase transition in the self-accelerating branch happens earlier than the transition in the normal branch. \\
In this case, our results could be compared with \cite{Suhonen,24,25,14ad,27,27a,27aa,Tawfik}. The time period of phase transition that was obtained in \cite{Suhonen,Tawfik}, where the first-order phase transition was studied in standard gravity model, is comparable with our result. By choosing a proper value of BD coupling constant, e.g. $\omega_c$, for our model, we could have almost the same period of time for the phase transition as in \cite{Suhonen,Tawfik}. However, for smaller values, $\omega < \omega_c$, the phase transition in our model happens faster. In \cite{24}, the authors utilized a RS brane-world framework and considered their results for different values of brane tension. It is shown that phase transition occurs much earlier than our work. Also, they considered brane tension effect on the time of transition which results in a later transition for a larger value of brane tension. In \cite{25} the authors have considered the phase transition in Brans-Dicke brane gravity and their results are plotted for different values of Brans-Dicke coupling constant, which displays that transition time increases by decreasing of BD coupling constant. The authors have considered the phase transition using a DGP brane scenario and their results are depicted for different values of brane tension in \cite{14ad} expressing later transition for a smaller value of brane tension. In comparison to \cite{25,14ad}, our model predicts that the quark-hadron phase transition lasts longer. \\
In the next part, the QCD phase transition was studied by applying a smooth crossover approach for two regimes as high and low temperatures. In the high-temperature regime, where the temperature is assumed to be bigger than $250 {\rm MeV}$, the trace anomaly is computed accurately and the outcome determines radiation like behavior for quarks and gluons. However, in the low-temperature regime, where the temperature is supposed to be less than $150 {\rm MeV}$, the trace anomaly is not applicable since it is affected by large discretization effect. Instead, the HRG model is utilized to construct a realistic equation of state. The time of phase transition in the high-temperature regime is the same for both normal and self-accelerating branches of the model. However, for the low temperature regime, the transition occurs faster for self-accelerating branch. The results of this case could be compared with \cite{14ad, 25, 27,27a,27aa}. In \cite{25}, the authors investigated QHP transition in RS brane-world model including a BD scalar field and in \cite{14ad} the topic is considered for a self-accelerating branch of DGP brane scenario. They found out a transition occurs in a few micro-second for both regimes. In \cite{27,27a,27aa}, the authors studied the phase transition in RS brane-world including a bulk chameleon-like scalar field. This model provides a non-conservation equation of state and their results express a phase transition happens faster than ours.

\section{Acknowledgments}
The work of T. G. has been supported financially by Vice Chancellorship of Research and Technology, University of Kurdistan under research Project No. 98/11/2724.
The work of A. M. has been supported financially by Vice Chancellorship of Research and Technology, University of Kurdistan under research Project No. 98/10/34704.

\vspace{1.5cm}


\end{multicols}

\clearpage
\end{CJK*}
\end{document}